\documentclass[twocolumn,amsmath,amssymb,pre]{revtex4}

\usepackage[]{graphicx} 
\usepackage{amsmath}

\begin{document}

\title{Coarse-Grained Counterions in the Strong Coupling Limit}

\author{Hiroshi Frusawa}
\email{frusawa.hiroshi@kochi-tech.ac.jp}

\affiliation{Laboratory of Statistical Physics, Kochi University of Technology, Tosa-Yamada, Kochi 782-8502, Japan.}

\date{\today}
\begin{abstract}
Recent Monte Carlo simulations (A. G. Moreira and R. R. Netz: Eur. Phys. J. E {\bf 8} (2002) 33) in the strong Coulomb coupling regime suggest strange counterion electrostatics unlike the Poisson-Boltzmann picture: when counterion-counterion repulsive interactions are much larger than counterion--macroion attraction, the coarse-grained counterion distribution around a macroion is determined only by the latter, and the former is irrelevant. Here, we offer an explanation for the apparently paradoxical electrostatics by mathematically manipulating the strong coupling limit.
\end{abstract}

\maketitle

Let us consider different types of materials, for instance, silica particles, DNA fragments, and clay. These have distinct shapes as well as chemical compositions: sphere, rod, and plate, respectively. Are there any similarities? To answer this positively, we need to put the materials in water where all of them become readily dispersed. Observing the suspensions, we can find high asymmetries of size and charge between dissociated counterions and nano-- to microscale macroions \cite{Israel, Safran, DNA, Andelman, soft}, which are in contrast to symmetric electrolytes (e.g., salty water). As a consequence of the asymmetries, some counterions are bound around macroions electrostatically and form an ionic cluster.

It has been known that the ionic cluster profile affects physical properties significantly \cite{Israel, Safran, DNA, Andelman, soft}. Many studies have thus addressed for decades the theoretical descriptions of counterion distribution around macroions \cite{Israel, Safran, DNA, Andelman, soft, review1, review2, others, Law, Moreira, Netz, Shklovskii}. From extensive theories, we can see two steps and two ways common to a rich variety of approaches. First, "two steps" imply mean-field approximations and advanced treatments. On the other hand, "two ways" indicate that mean-field approximations can be made in both extremes of weak and strong Coulomb coupling specified later.

As a matter of fact, however, most theories \cite{soft, review1, review2, others} have started with the Poisson-Boltzmann (PB) approaches valid in the weak coupling regime. It is only in recent years that a few groups \cite{Law, Moreira, Netz, Shklovskii} have considered strong coupling approximations.

Why has minimal attention been paid to the strong coupling approach? One of the reasons is the ambiguity of the Coulomb coupling for counterions. Since we have had no consensus on the definition of either strong or weak Coulomb coupling, it has not been discriminated clearly whether the present simulations or experiments are located in the strong coupling regime.

This situation differs from that for symmetric electrolytes (or plasmas) which have an appropriate measure. The standard is called the coupling constant defined as $\Gamma = z^2l_B/a$ \cite{plasma}. Here, $z$ is the valence of charges, $a$ is the Wigner-Seitz cell size, and $l_B\equiv e^2/4\pi\epsilon k_BT$ is the distance (the so-called Bjerrum length) at which two elementary charges interact electrostatically with thermal energy $k_BT$ when they are surrounded by the polar solvent with its dielectric permittivity and temperature being $\epsilon$ and $T$. By the coupling constant $\Gamma$, the weak and strong coupling regimes for plasmas have been represented as $\Gamma <<1$ and $\Gamma >>1$, respectively. For counterions, however, it is inappropriate to take the Wigner-Seitz cell size $a$ in rescaling the Bjerrum length as $l_B/a$, because counterions under macroion potential cannot form a uniform Wigner crystal.

Recently, Moreira and Netz have proposed a new coupling constant for counterions \cite{Moreira}: $\Gamma = z^2l_B/\lambda$, by introducing the Gouy-Chapman length $\lambda=1/(2\pi zl_B\sigma)$ as an alternative length, where $z$ is now the valence of counterions and $\sigma$ the surface number density of macroion charges. Using the new coupling constant $\Gamma = z^2l_B/\lambda=z^3\sigma l_B^2$, they classified the results of Monte Carlo simulations for the counterion distribution around planar macroions, and revealed novel counterion electrostatics in the strong Coulomb coupling regime ($\Gamma=z^2l_B/\lambda >>1$) as follows \cite{Moreira}: only the bare (or unscreened) potential of charged plate(s) determines the counterion distribution over a wide range far beyond the Gouy-Chapman length $\lambda$, when the valence $z$ is multivalent, and macroion charge density $\sigma$ is high to some extent, and temperature $T$ and/or permittivity $\epsilon$ of the solvent is lowest (namely, $\Gamma>>1$).

To be more specific for a single-charged plate, their simulations show that the counterion density $C_{\rm SC}(h)$ as a function of height $h$ from the plate is not an algebraically decaying profile of the PB solution, but becomes an exponentially decaying one: the Boltzmann distribution $C_{\rm SC}(h)\propto \exp(-h/\lambda)$ with the weight of macroion potential ($\sim h/\lambda$) is valid even for $h/\lambda>>1$. The result explicitly contradicts the PB picture that the potential which counterions feel is screened by counterions themselves self-consistently.

To explain the novel counterion electrostatics, it seems natural to seek a strong coupling approach without resorting to the weak coupling theories which incorporate counterion fluctuations and correlations into the PB-type approximations. 

Indeed strong coupling approaches \cite{Moreira, Netz, Shklovskii} have been already presented, as mentioned above. While one justifies the use of the virial expansion via field theoretic formulations \cite{Moreira, Netz}, the other supposes that all counterions should condense on the macroion surface to form a 2D Wigner crystal in the ground state (i.e., $\Gamma\rightarrow\infty$) \cite{Shklovskii}. Both explain the above counterion distribution, $C_{\rm SC}(h)\propto \exp(-h/\lambda)$: one \cite{Moreira, Netz} by the leading order (no interaction term) of the virial expansion, and the other \cite{Shklovskii} by an intuitive discussion which focuses on one counterion escaping from the macroion surface. Despite the preceding considerations \cite{Moreira, Netz, Shklovskii}, however, it remains an open problem why counterion-counterion repulsions have no effect on the primary counterion distribution around a macroion in the strong coupling limit; at first glance, this is paradoxical.

Hence, this letter aims to solve the apparent paradox in terms of a functional integral form applicable to any shapes of macroions. A key procedure is to introduce Coulomb potential fields twice: we insert a real potential field into the conventional field theoretic formulations for imaginary potential and density variables \cite{Orland}. The technical roundabout way enables us to provide a clearer view of the strong coupling system than previous explanations.

\begin{figure}[htbp]
	\begin{center}
	\includegraphics[width=7.8cm]{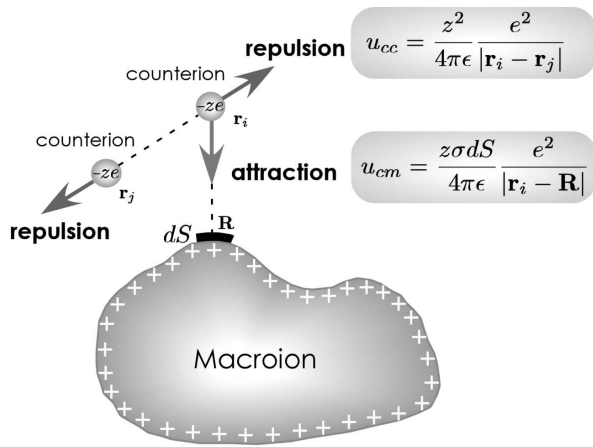}
	\end{center}
	\caption{Schematic of the present system consisting of one fixed macroion with positive charges and oppositely charged counterions. Both counterion-counterion repulsion and counterion-macroion attraction act on a counterion. Also, two energies, $u_{cc}$ and $u_{cm}$, are written down to illustrate the notation in the text, where $u_{cc}$ denotes a repulsive energy between counterions at the positions ${\bf r}_i$ and ${\bf r}_j$, and $-u_{cm}$, an attractive one between a counterion at ${\bf r}_i$ and a surface element with its area $dS$ at the position ${\bf R}$. In the expression of $u_{cm}$, $e\sigma\,dS$ is the total charge of the surface element.
	}
	\label{geometry}
\end{figure}

{\itshape Setting up the Problem}--- Figure 1 shows the geometry and notation for the problem. We consider the system, with its volume $\Omega$, which consists of a fixed macroion positively charged and $N$--counterions with opposite charges of $z$--valence. All of the discussions presented herein can be extended straightforwardly to multimacroion systems. The counterion number $N$ is imposed on the global electroneutrality condition, $-zN+\sigma\Sigma=0$ ($\Sigma$: the surface area of a macroion). The system has electrostatic interaction energy $E_{cc}+E_{cm}$, i.e., the sum of the counterion-counterion interaction energy $E_{cc}$ and the counterion-macroion interaction energy $E_{cm}$. In the thermal energy unit, we have
    \begin{eqnarray}
    E_{cc}
    &=&\frac{z^2\,l_B}{2}\left(
    \sum_{i,j=1}^N\,v({\bf r}_i-{\bf r}_j)-N\,v(0)
    \right)\nonumber\\
    E_{cm}
    &=&-z\,l_B\,\sigma\sum_{i=1}^N\,\oint\,dS\>v({\bf r}_i-{\bf R}),
    \label{energy}
    \end{eqnarray}
where $v({\bf r})\equiv1/|{\bf r}|$, and $\oint\,dS$ denotes that the integration of the macroion charge position ${\bf R}$ is restricted to the surface of a macroion.

{\itshape Our Scenario}--- According to the PB theory, the counterion distribution $C_{\rm{PB}}({\bf r})$ around a macroion is expressed as
   \begin{eqnarray}
   C_{\rm{PB}}({\bf r})
   &=&N\frac{\exp\,[-\psi_c^{\rm{PB}}({\bf r})-\psi_m({\bf r})\,]}
   {\int\,d{\bf r}\>\exp\,[-\psi_c^{\rm{PB}}({\bf r})-\psi_m({\bf r})\,]}
   \label{pb},
   \end{eqnarray}
where $\psi_c^{{\rm PB}}$ and $\psi_m$ are the products of $ze/k_BT$ multiplied by the electrostatic potentials due to counterions and macroion, respectively. While the counterion potential $\psi_c({\bf r}_i)=z^2l_B\sum_{j=1}^N v({\bf r}_i-{\bf r}_j)$ is replaced by the mean-field one $\psi_c^{{\rm PB}}$ obtained from solving the PB equation, the macroion potential $\psi_m$ has been set as $\psi_m({\bf r})=-zl_B\,\sigma\,\oint\,dS\>v({\bf r}-{\bf R})$.

In the strong coupling limit, on the other hand, the Monte Carlo simulations by Moreira and Netz \cite{Moreira} suggest the counterion distribution $C_{\rm{SC}}({\bf r})$ as follows:
  \begin{eqnarray}
   C_{\rm{SC}}({\bf r})
   =N\frac{\exp\,[-\psi_m({\bf r})]}
   {\int\,d{\bf r}\>\exp\,[-\psi_m({\bf r})]}.
   \label{sc_density}
   \end{eqnarray}
Comparing eq. (\ref{sc_density}) with eq. (\ref{pb}), it naturally occurs to us that eq. (\ref{pb}) reduces to eq. (\ref{sc_density}) when the counterion potential $\psi_c$ is a spatially independent constant. If so, a plausible selection of the constant would be $\psi_c^0=z^2l_B\overline{\rho}\int d{\bf r}\;v({\bf r})$, where $\overline{\rho}=N/\Omega$ is the counterion density smeared uniformly over the  entire system. This constant potential $\psi_c^0$ corresponds to the zero--wave--number quantity, $\psi_c({\bf k}\rightarrow 0)$, of the above counterion potential $\psi_c$ which is written in the Fourier-transformed representation as $\psi_c({\bf k})=z^2l_B\hat{\rho}({\bf k})\,(4\pi/{\bf k}^2)$ with using $\hat{\rho}({\bf k})=\Omega^{-1}\sum_{j=1}^N\exp(i{\bf k}\cdot{\bf r}_j)$. It does not matter here if the constant potential $\psi_c^0=\psi_c({\bf k}\rightarrow 0)$ is infrared-divergent because this constant term is cancelled in eq. (\ref{sc_density}) between the denominator and numerator and is irrelevant in the result.

To summarize, our scenario assumes that the counterion potential $\psi_c$ reduces to a constant:
   \begin{equation}
   \psi_c({\bf r})\rightarrow \psi_c^0
   \label{proof}
   \end{equation}
in the strong coupling limit, $\Gamma \rightarrow \infty$, quite differently from the weak coupling approximations where $\psi_c\rightarrow \psi_c^{\rm PB}$. Can we offer a simple explanation for the scenario?

{\itshape Underlying Physics}--- To consider the physics behind our speculation, let us have preliminary discussions before tracing the mathematics. We first rescale the system as $\widetilde{\bf r}={\bf r}/\lambda$, and convert the energy forms (\ref{energy}) into
   \begin{eqnarray}
    E_{cc}
    &=&\frac{\Gamma}{2}\left(
    \sum_{i,j=1}^N\,v(\widetilde{\bf r}_i-\widetilde{\bf r}_j)
    -Nv(0)
    \right)
    \nonumber\\
    E_{cm}
    &=&-\frac{1}{2\pi}\sum_{i=1}^N\,\oint\,d\widetilde{S}\>
    v(\widetilde{\bf r}_i-\widetilde{\bf R}),
    \label{rescale-energy}
    \end{eqnarray}
where we have used the relation $d\widetilde{S}=dS/\lambda^2$ and the coupling constant $\Gamma=z^2l_B/\lambda$ proposed by Moreira and Netz. The rescaled expressions (\ref{rescale-energy}) reveal that the strong coupling condition $\Gamma>>1$ leads to the extreme asymmetry $E_{cc}>>E_{cm}$. Consequently, in the strong coupling limit, while the counterion-macroion interaction energy $E_{cm}$ is finite, the counterion-counterion interaction energy $E_{cc}$ approaches infinity even after the infrared- and ultraviolet--divergent terms are regularized. From this we find that, in the strong coupling regime, the energetic gain due to counterion--macroion interactions is not essential, and minimizing counterion--counterion interactions should be carried out first of all.

Nonetheless, the previous discussions \cite{Shklovskii} based on the 2D plasma theory take it for granted that the localization of counterions on the macroion surface is the ground state and minimizes the total energy; actually, however, since the complete fusion of counterions and macroion surface, giving no total charge, is never allowed, the counterion localization has much energy requirement due to the predominant repulsion energy $E_{cc}$.

To obtain the correct ground state, it is necessary to rewrite $E_{cc}$ in the potential form as $E_{cc}=(8\pi\Gamma)^{-1}\int d{\bf r} |\nabla\psi_c({\bf r})|^2+E_{cc}^0$, where $E_{cc}^0$ is an offset energy given by $E_{cc}^0=N/2[\psi_c^0-v(0)]$. We immediately find from the expression that the lowest bound $E_{cc}\rightarrow E_{cc}^0$ is realized for $|\nabla\psi_c|=0$, i.e., a spatially invariant potential $\psi_c^0$, in agreement with the above supposition (\ref{proof}). In other words, the strong coupling systems are stabilized when varying the counterion potential $\psi_c$ causes little change in the counterion--counterion interaction energy $E_{cc}$, which results in the present distribution (\ref{sc_density}).

The remainder of this letter will now be devoted to validating the relation (\ref{proof}) from the point of functional integral view; the mathematical manipulations given below are truly our result.

{\itshape Conventional Start}--- To obtain the average counterion density, we define the canonical partition function $Z\{J\}={\rm Tr}\>\exp\,[-E_{cc}-E_{cm}+\sum_{i=1}^N J(\widetilde{\bf r}_i)]$. Here, the operator "${\rm Tr}$" reads ${\rm Tr}\equiv (1/N!)\,\prod_{i=1}^N\,\int\,d{\bf r}_i$ while setting the thermal wavelength to unity for simplicity, and the external source $J$ is a probe for averaging the instantaneous density $\hat{\rho}({\bf r})$: $C({\bf r})\equiv<\widehat{\rho}({\bf r})>=\delta(\ln Z)/\delta J|_{J=0}$ which is the average density in the absence of the external source.
  
Inserting into $Z\{J\}$, as usual, the trivial constant $1=\int D\rho\>\prod_{\{\bf r\}}\delta[\widehat{\rho}({\bf r})-\rho({\bf r})]$ and exponentiating the operation of ${\rm Tr}$, the partition function reads
    \begin{eqnarray}
    &&Z=\int D\rho\>D\phi\;e^{-\mathcal{F}\{\rho,\phi,J\}}
    \nonumber\\
    &&\mathcal{F}\{\rho,\phi,J\}
    =\frac{\Gamma}{2}\int d{\bf r}d{\bf r}'\>
    \rho({\bf \widetilde{r}})v({\bf \widetilde{r}}-{\bf \widetilde{r}}')
    \rho({\bf \widetilde{r}}')\nonumber\\
    &&\qquad\quad+\int d{\bf \widetilde{r}}\;
    i\rho({\bf \widetilde{r}})\phi({\bf \widetilde{r}})
    +N\mu\{\phi,J\}-N,
    \label{convention}
    \end{eqnarray}
where subtraction of the self-energy is not written explicitly for brevity, and $\mu\{\phi,J\}$ corresponds to chemical potential of counterions in terms of the grand canonical system: $\mu\{\phi,J\}=\ln\,\left[N/\int d\widetilde{\bf r}\> e^{i\phi-\psi_m+J}\right]$ giving $\mu\{0,0\}=\ln\overline{\rho}$ in ideal systems. The canonical expression (\ref{convention}) indicates that the chemical potential term (or the fugacity one) is independent of the coupling constant $\Gamma$ and therefore does not become perturbative even in the limit $\Gamma\rightarrow\infty$, although the $\Gamma$--dependence has been ambiguous in the grand canonical form \cite{Netz}. That is, there is no proper basis for the fugacity expansion (or the virial one) in the strong coupling regime.
  
Furthermore, performing Gaussian integration over the $\rho$--field in eq. (\ref{convention}), we obtain the conventional field-theoretic form for the potential field $\{\phi\}$, which is called the sine-Gordon theory in the case of symmetric electrolytes \cite{Orland}. As described above, however, we eliminate the $\rho$--field in the Hamiltonian $\mathcal{F}\{\rho,\phi,J\}$, not by the Gaussian integration but by introducing another field of electrostatic potential; otherwise, the conventional formulation in the canonical system leads to the average density whose form is, in the strong coupling limit,  mathematically complicated and physically unclear \cite{bussei}.

{\itshape Adding Potential Field Once More}--- We would like to introduce the real potential $\psi$ (in the unit of $k_BT/ze$ as before), simply via the Poisson equation $\nabla^2\psi({\bf \widetilde{r}})=-4\pi\Gamma(\rho({\bf \widetilde{r}})-\overline{\rho})$ in the rescaled system, which implies $\psi({\bf r})=\Gamma\int d{\bf \widetilde{r}'\>(\rho({\bf \widetilde{r}'})-\overline{\rho})\>v({\bf \widetilde{r}}-{\bf \widetilde{r}'})}$. To this end, we insert into eq. (\ref{convention}) a trivial integral, $1=\mathcal{N}^{\,2}\int D\psi\;\,\Delta\{\rho,\psi\}$, where $\mathcal{N}^{\,2}={\rm Det}[-\nabla^2/(4\pi\Gamma)]$, and $\Delta\{\rho,\psi\}$ is a $\delta$-functional given by $\Delta\{\rho,\psi\}=\prod_{\{\bf \widetilde{r}\}}\>\delta\left[-\nabla^2\psi({\bf \widetilde{r}})/(4\pi\Gamma)-\left(\rho({\bf \widetilde{r}})-\overline{\rho}\right)\right]$. Equation (\ref{convention}) then reads
   \begin{eqnarray}
    &&Z=\mathcal{N}^{\,2}
    \int D\rho\>D\psi\>D\phi\;\>
    \Delta\{\rho,\psi\}\>
    e^{-\mathcal{F}\{\rho,\psi,\phi,J\}}
    \nonumber\\
    &&\mathcal{F}\{\rho,\psi,\phi,J\}
    =\frac{1}{8\pi\Gamma}\int d{\bf r}\>
    \left|
    \nabla\psi(\bf \widetilde{r})
    \right|^2
    -\frac{N}{2}\psi_c^0\nonumber\\
    &&+\int d{\bf \widetilde{r}}\;i\rho({\bf \widetilde{r}})
    \left[
    \phi({\bf \widetilde{r}})
    -i\psi_c^0
    \right]
    +N\mu\{\phi,J\}-N.
    \label{new_form}
    \end{eqnarray}
Here, in using the relation $\psi(-\nabla^2)\psi=-\nabla\cdot(\psi\nabla\psi)+\nabla\psi\cdot\nabla\psi$, we have considered that the surface integral term vanishes due to the definition of $\psi$.

The novel expression (\ref{new_form}) of Coulombic systems is relevant to the strong coupling regime as seen below. When we carry out the functional integral over the $\rho$-- and $\psi$-- fields while recovering the relation that $\rho(\widetilde{\bf r})-\overline{\rho}=-\nabla^2\psi(\widetilde{\bf r})/4\pi\Gamma$ in eq. (\ref{new_form}), we can check that our form (\ref{new_form}) is consistent with the conventional sine-Gordon theory; the normalization factor is also reduced to the familiar one [$\sim{\rm Det}^{1/2}(-\nabla^2/\Gamma)$] due to the Gaussian integration over the $\psi$-field.

{\itshape Strong Coupling Approximation}--- Since the first electrostatic energy term on the right--hand side of eq. (\ref{new_form}) is negligible in the strong coupling limit $\Gamma\rightarrow\infty$ as deliberated below, the functional integral over the $\psi$--field reduces to the trivial one, $1=\mathcal{N}^{\,2}\int D\psi\;\,\Delta\{\rho,\psi\}$. In this case, the functional integral over the $\rho$--field yields the following $\delta$-functional:
    \begin{eqnarray}
    &&Z\approx
    \int D\phi\;\>\prod_{\{\bf \widetilde{r}\}}
    \delta[\phi({\bf \widetilde{r}})-i\psi_c^0]
    \;\>e^{-\mathcal{F}\{\phi,J\}}
    \nonumber\\
    &&\mathcal{F}\{\phi,J\}
    =
    N\mu\{\phi,J\}-N-\frac{N}{2}\psi_c^0,
    \label{sc}
    \end{eqnarray}
which supports our central relation (\ref{proof}) in the strong coupling limit, because we can regard $-i\phi$ as a physically relevant potential (see the $\mu$--form). It is straightforward to confirm that the approximate partition function (\ref{sc}) gives the average density $C_{\mathrm{SC}}({\bf r})=\delta(\ln Z)/\delta J|_{J=0}$ as
   \begin{eqnarray}
   C_{\mathrm{SC}}({\bf r})&=&-\frac{N}{\lambda^3}\left.
  \frac{\delta\mu\{i\psi_c^0,J\}}{\delta J(\widetilde{\bf r})}
  \right|_{J=0},
  \label{sc_result}
  \end{eqnarray}
which reduces to eq. (\ref{sc_density}) because the counterion potential with a constant $\psi_c^0$ cancels between the denominator and numerator, as speculated in "{\itshape Our Scenario}".

To clarify the physical meaning of the approximate form (\ref{sc}) more explicitly, it is appropriate to express eq. (\ref{sc}) in the form of the Helmholtz free energy $A=-\ln Z$:
    \begin{eqnarray}
    A&=&E_{cc}^0+\int d{\bf r}\;
    C_{\mathrm{SC}}({\bf r})\psi_m({\bf r})\nonumber\\
    &&\quad+C_{\mathrm{SC}}({\bf r})\ln C_{\mathrm{SC}}({\bf r})
    -C_{\mathrm{SC}}({\bf r}),
    \label{helm}
    \end{eqnarray}
using the average concentration $C_{\mathrm{SC}}({\bf r})$ given by (\ref{sc_density}). The first line is the energetic term, and the second, the contribution of counterion entropy. Here, we would like to repeat the difference from the PB expression: the counterion--counterion energy $E_{cc}$, divergent in the strong coupling limit, is suppressed to have an offset energy $E_{cc}^0$.

{\itshape Connection with Coarse-Graining}--- To be precise, coarse-graining is implicit in the strong coupling approximation, as found from the Fourier-transformed representation of the first term of eq. (\ref{new_form}): $1/(8\pi\Gamma)\int d\widetilde{\bf r}\,|\nabla\psi(\widetilde{\bf r})|^2=\sum_k{\widetilde{\bf k}}^2/(8\pi\Gamma)\,\psi_{\widetilde{\bf k}}^2$. The above wave--vector--dependent expression indicates that, for the above approximation, a cutoff length $\widetilde{a}_c=2\pi/|{\bf \widetilde{k}}_c|$ is required to satisfy $\widetilde{a}_c>>\Gamma^{-1/2}$. Is it possible to really set the cutoff length fulfilling the criterion?

Let us then consider the shortest possible cutoff length, or the average separation between counterions when they seem most crowded. A plausible situation is that all counterions condense on the macroion surface, where they are at a distance [$\sim(z/\sigma)^{1/2}$] \cite{Shklovskii}. When we ignore counterion--counterion correlations below this scale and set $a_c\sim(z/\sigma)^{1/2}$, we obtain the relation $(a_c/\lambda)\Gamma^{1/2}\sim\Gamma>>1$, or $\widetilde{a}_c=a_c/\lambda>>\Gamma^{-1/2}$, implying that even the shortest cutoff length (or the largest cutoff wave number) can satisfy the coarse-graining criterion for our approximation.


{\itshape Concluding Remarks}--- We have thus confirmed, in a field-theoretic way, that the novel relation (\ref{proof}) explains the counterion distribution (\ref{sc_density}) suggested by recent Monte Carlo simulations in the strong coupling regime \cite{Moreira}. Through the derivations, logical skips in the previous approaches \cite{Netz, Shklovskii} have also been disclosed: our framework does not justify their starting points, i.e., virial expansion \cite{Netz} and 2D Wigner crystal \cite{Shklovskii}. Without the preceding considerations, however, a simple picture represented by eqs. (\ref{proof}) and (\ref{helm}) is now available: since the sum of counterion-counterion interactions becomes insensitive to counterion arrangements in the strong coupling limit, the counterion distribution is determined only by the counterion-macroion interactions.


\end{document}